\documentclass[fleqn,twoside]{article}
\usepackage[headings]{espcrc2}

\readRCS
$Id: espcrc2.tex,v 1.2 2004/02/24 11:22:11 spepping Exp $
\usepackage{graphicx}
\usepackage[figuresright]{rotating}

\newcommand{\AmS}{{\protect\the\textfont2
  A\kern-.1667em\lower.5ex\hbox{M}\kern-.125emS}}

\title{Irrational constants in positronium decays}

\author{B.  A. Kniehl,\address[II]{II Institut f\"ur Theoretische Physik, 
        Universit\"at Hamburg\\
        22761 Hamburg, Germany}
        A. V. Kotikov,\addressmark[II]
        \address{
        Bogolyubov Laboratory for Theoretical Physics, JINR\\
        141980 Dubna, Moskau region, Russia
        }
        \underline{O. L. Veretin}\addressmark[II]
        \address{
        On leave of absence from\\
        Petrozavodsk State University,\\
        185910 Petrozavodsk, Karelia, Russia
        }
        \thanks{
        This work was supported in part by BMBF Grant No.\ 05~HT6GUA, DFG Grant No.\
        SFB~676, and HGF Grant No.\ NG-VH-008.
        }
}       

\runtitle{2-column format camera-ready paper in \LaTeX}
\runauthor{O. L. Veretin}

\begin{document}

\begin{abstract}
We establish irrational constants, that contribute to the positronium lifetime at
$O(\alpha)$ and $O(\alpha^2)$ order. In particular we show, that a new type
of constants appear, which are not related to Euler--Zagier sums or multiple
$\zeta$ values.
\end{abstract}

\maketitle

\section{Introduction}

Most of the multi-loop analytical calculations in quantum field theories
have been dine for so-called single-scale problems. This means that the
evaluated integrals are basicaly expressed as numerical constants 
up to a trivial scale factor. Examples of such problems include
almost all renormalization group calculations, evaluations of the
critical exponents, anomalous magnetic moments of the electron and the muon,
matching calculations in effective theories (e.g. HQFT, NRQFT) and many others.

  Usually analytical results involve the so-called Euler--Zagier (EZ) sums 
of the form
\begin{equation}
  \sum_{n_1>n_2>\dots>n_k}
     \frac{(\pm1)^{n_1}}{n_1^{a_1}}
     \dots
     \frac{(\pm k)^{n_k}}{n_1^{a_k}}
\end{equation}
or more generally multiple polylogarithms
\begin{equation}
  \sum_{n_1>n_2>\dots>n_k}
     \frac{z_1^{n_1}}{n_1^{a_1}}
     \dots
     \frac{z_k^{n_k}}{n_1^{a_k}}
\end{equation}
where $z_1,\,\dots z_k$ are some parameters and $a_1,\,\dots,a_k$
are positive integers. The sum $a_1 + a_2 + \dots + a_k$
is called the {\it weight} in such a case.

   The above definitions include e.g. well-known\\ irrationalities like
$\zeta$ functions $\zeta(a),\,\zeta(a,b),\,\dots$, (poly)logarithms
${\rm Li}_a(1/2),\,\ln 2,\,\dots$ and ``sixth\\ root of unity'' constants
${\rm Ls}_j^{(k)}(\pi/3),\,{\rm Ls}_j^{(k)}(2\pi/3),\\\,\dots$.
There is no doubt, that by consideration of more complicated problems
and in higher loops some mew constants will appear.
As examples we can mention some elliptic integrals (see e.g. \cite{I1,I2,I3}).

 In this paper we concentrate on a very important single-scale problem:
the total width of positronium decay in QED. Positronium (Ps),
the lightest known atom, provides an ultra-pure laboratory for 
high-precision tests of QED. In fact, thanks to the smallness of the
electron mass $m$ relative to typical hadronic mass scale, its
theoretical description is not plagued by strong interaction uncertainties
and its properties, such as decay widths and energy levels 
can be calculated perturbatively in non-relativistic QED (NRQED) \cite{Caswell:1985ui}
with very high precision.

  Ps comes in two ground states, ${}^1S_0$ parapositronium ($p$-Ps)
and ${}^3S_1$ orthopositronium ($o$-Ps), which decay to two and three
photons, respectively.

\section{Orthopositronium}

  In this section we are concerned with the lifetime of $o$-Ps,
which has been the subject of a vast number of theoretical and
experimental investigations. Its first precision measurement \cite{BH},
of 1968, had to wait nine years to be compared with first complete
one-loop calculation \cite{Caswell:1976nx}, which came two decades after the
analogous calculation for $p$-Ps \cite{HB} being considerably simpler
owing to the two-body final state. In the year 1987, the Ann Arbor 
group \cite{I8} published a measurement that exceeded the theoretical 
prediction avalaible by ten experimental stantard deviations.
This is so-called $o$-Ps lifetime puzzule triggered an avalanche of
both experimental and theoretical activities, which eventually 
resulted in what now appears to be the resolution of this puzzle.
In fact, the 2003 measurements at Ann Arbor \cite{I9} and
Tokio \cite{I10}
\begin{eqnarray}
\Gamma(\mbox{Ann Arbor}) &=&
7.0404(10)(8)~\mu s^{-1},
\nonumber\\
\Gamma(\mbox{Tokyo}) &=&
7.0396(12) (11)~\mu s^{-1},  
\end{eqnarray}
agree mutually and with the present theoretical prediction,
\begin{equation}
\Gamma(\mbox{theory}) = 7.039979(11)~\mu s^{-1}.
\end{equation}
   The latter is evaluated from
\begin{eqnarray}
\Gamma(\mbox{theory}) &=& \Gamma_0\left[1 + A \frac{\alpha}{\pi}
+\frac{\alpha^2}{3} \ln\alpha \right.
\nonumber\\
 &+& B \left(\frac{\alpha}{\pi}\right)^2 
   - \frac{3\alpha^3}{2\pi} \ln^2 \alpha 
\nonumber\\
  &+& \left. C \frac{\alpha^3}{\pi} \ln \alpha   \right],
\label{Gamma}
\end{eqnarray}
where \cite{Ore:1949te}
\begin{equation}
\Gamma_0 = \frac{2}{9}(\pi^2-9)\frac{m\alpha^6}{\pi}
\end{equation}
is the LO result.
The leading logarithmically enhanced ${\mathcal O}(\alpha^2\ln\alpha)$ and
${\mathcal O}(\alpha^3\ln^2\alpha)$ terms were found in
Refs.~\cite{Caswell:1978vz,Khriplovich:1990eh} and Ref.~\cite{Kar},
respectively.
The coefficients $A=-10.286606(10)$
\cite{Caswell:1976nx,Caswell:1978vz,SH,Adkins:2000fg,Adkins:2005eg},
$B=45.06(26)$ \cite{Adkins:2000fg}, and $C=-5.51702455(23)$
\cite{Kniehl:2000dh} are only available in numerical form so far.
Comprehensive reviews of the present experimental and theoretical status of
$o$-Ps may be found in Ref.~\cite{AFS}.

Given the fundamental importance of Ps for atomic and particle physics, it is
desirable to complete our knowledge of the QED prediction in
Eq.~(\ref{Gamma}).
Since the theoretical uncertainty is presently dominated by the errors in the
numerical evaluations of the coefficients $A$, $B$, and $C$, it is an urgent
task to find them in analytical form, in terms of irrational numbers,
which can be evaluated with arbitrary precision.
In this Letter, this is achieved for $A$ and $C$.
The case of $B$ is beyond the scope of presently available technology, since
it involves two-loop five-point functions to be integrated over a three-body
phase space. 
The quest for an analytic expression for $A$ is a topic of old vintage:
about 25 years ago, some of the simpler contributions to $A$, due to
self-energy and outer and inner vertex corrections, were obtained analytically
\cite{Stroscio:1982wj}, but further progress then soon came to a grinding halt.

\begin{figure}[ht]
\begin{center}
\includegraphics[width=0.45\textwidth]{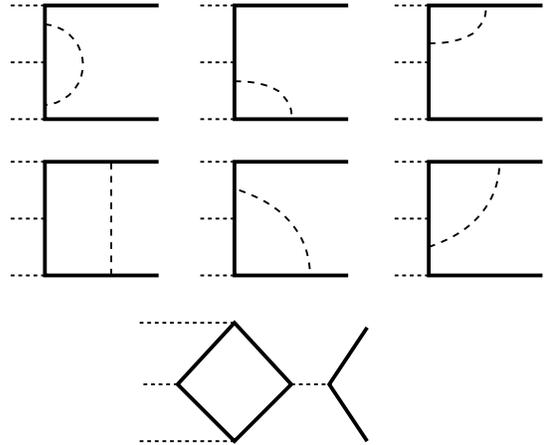}
\end{center}
\caption{
Feynman diagrams contributing to the total decay width of $o$-Ps at
${\mathcal O}(\alpha)$.
Self-energy diagrams are not shown.
Dashed and solid lines represent photons and electrons, respectively.}
\end{figure}
The ${\mathcal O}(\alpha)$ contribution in Eq.~(\ref{Gamma}),
$\Gamma_1=\Gamma_0A\alpha/\pi$, is due to the Feynman diagrams where a virtual
photon is attached in all possible ways to the tree-level diagrams, with three
real photons linked to an open electron line with threshold kinematics.
Such diagrams are shown in Fig.~1.

  After angular integration over three-photon phase space
\begin{equation}
  \int [dk_1] [dk_2] [dk_3] \delta(k_1+k_2+k_3-q)
\end{equation}
we can rewrite the one-loop contribution to the width as (see \cite{Adkins:2005eg})
\begin{eqnarray}
\Gamma_1  &=& \frac{m\alpha^7}{36\pi^2}
\int\limits^1_0\frac{{\mathrm d}x_1}{x_1}\,\frac{{\mathrm d}x_2}{x_2}\,
\frac{{\mathrm d}x_3}{x_3}\delta(2-x_1-x_2-x_3)
\nonumber\\
 && {}\times[F(x_1,x_3) + {\mathrm{perm.}}],
\label{eq:org}
\end{eqnarray}
where $x_i$, with $0\le x_i\le 1$, is the energy of photon $i$ in the $o$-Ps
rest frame normalized by its maximum value, the delta function ensures energy
conservation, and {\it perm.}\ stands for the other five permutations of
$x_1,x_2,x_3$.

  The function $F$ includes dilogarithm and arc\-tangent functions as given in \cite{Adkins:2005eg}.
Ror illustration, we just mention, that the above expression, after re-parametrization,
consists of integrals of the following type
\begin{eqnarray}
  \frac{P(x_1,x_2,x_3)}{Q(x_1,x_2,x_3)}
  \int\limits_0^1 \frac{dy\, \ln(x_1+(1-x_1)y^2)}{(1-x_1)x_3-x_1(1-x_3)y^2}\,,
  \nonumber\\
  \frac{P(x_1,x_2,x_3)}{Q(x_1,x_2,x_3)}
  \int\limits_0^1 \frac{dy\, \ln(x_1+(1-x_1)y^2)}{x_1 x_3-(1-x_1)(1-x_3)y^2}\,,
  \nonumber
\end{eqnarray}
with $P,Q,P'Q'$ being some polynomials.

  The analytical integration of the above expressions is rather tedious and 
requires a number of tricks, e.g. expansion in series. Only a few integrals
could be done strightforwardly, e.g.. with {\it Mathematica} or {\it Maple}.
However, we established all irrational constants in terms of which the complete
one-loop correction can be expressed. These include among others usual EZ sums
up to weigth four, including e.g.
\begin{eqnarray}
  \ln2 \,, \qquad \zeta(n) \,, \qquad {\rm Li}_4 \left( \frac{1}{2} \right)
   \,, \quad\mbox{etc.}  \nonumber
\end{eqnarray}
and some additional constants of new type. At weight one, we have
\begin{eqnarray}
 \ln(R) \,, \quad \mbox{where} \quad  R = \frac{\sqrt{2}-1}{\sqrt{2}+1} \nonumber
\end{eqnarray}
and up to weight four our basis includes the following constants
\begin{eqnarray}
&& {\rm Li}_2 \left( \frac{1}{3} \right) \,, \qquad
{\rm Li}_4 \left( \frac{1}{3} \right) \,, \qquad
{\rm Li}_4 \left( -\frac{1}{3} \right) \,, \nonumber \\
&& {\rm Li}_3 \left( \frac{1}{\sqrt{2}} \right) \,, \qquad
{\rm Li}_3 \left( R \right) \,, \qquad
{\rm S}_{1,2} \left( R \right) \,, \qquad \nonumber \\
&& {\rm Li}_4 \left( \pm R \right) \,, \qquad
{\rm S}_{1,3} \left( \pm R \right) \,, \qquad
{\rm S}_{2,2} \left( \pm R \right) \,,\nonumber
\end{eqnarray}
with ${\rm S}_{a,b}$ being the generalized polylogarithm
\begin{eqnarray}
{\rm S}_{a,b}(x) = \frac{(-1)^{a+b-1}}{(a-1)!b!} \int\limits_0^1
      \frac{dt}{t} \ln^{a-1}t\, \ln^b(1-tx) \,.\nonumber
\end{eqnarray}
Unfortunately, not all integrals can be computed analytically.
In more complicated cases, the integrations are not separated 
after expansion into infinite series. We then rely on the PSLQ
algorithm \cite{PSLQ}, which allows one to reconstruct the
representation of a numerical result known to very high precision
in terms of a linear combinations of a set of constants with
rational coefficients, if that set is known beforehand. The experience 
gained with the explicit solution of the simpler integrals helps us
to exhaust the relevent set. 
In order for PSLQ to work in our applications, the numerical values of the
integrals must be known up to typically 150 decimal figures.

\section{Parapositronium}

Let us now turn to the case of parapositronium. Its total width
was recently measured to be \cite{I28}
\begin{equation}
  \Gamma_p({\rm exp}) = 7990.9\mu s^{-1} \,.
\end{equation}

  At present, the following radiative corrections within NRQED are
available:
\begin{eqnarray}
&&
\Gamma_p ~=~ \frac{\alpha^5\,m_e}{2} \Biggl\{
     1
    + \frac{\alpha}{\pi}  \, \left(\frac{\pi^2-20}{4} \right)
\nonumber\\
&&
    + \frac{\alpha^2}{\pi^2}  \, \left( - 2\pi^2 \ln\alpha + A_p \right)
    + \frac{\alpha^3}{\pi} \left(
        - \frac{3}{2} \ln^2\alpha \right.
\nonumber\\
&&
       \left.
        + ( \frac{533}{90} - \frac{\pi^2}{2} + 10\ln2 ) \ln\alpha  \right)
     \Bigg\} \,.
  \nonumber
\end{eqnarray}
The first-order corrections were obtained in \cite{HB}, while the
logarithmically enhanced terms were computed in \cite{Caswell:1978vz,Khriplovich:1990eh}.
Here the constant $A_p=5.12443(33)$ is known only numerically \cite{I20}
and our next goal is to establish the irrational constants that contribute
to this quantity.

\begin{figure}[ht]
\begin{center}
\includegraphics[width=0.45\textwidth]{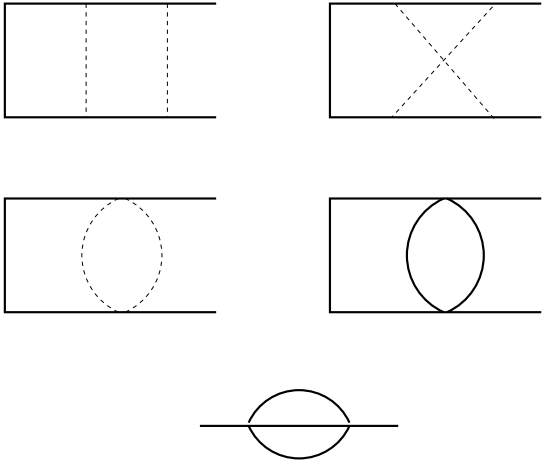}
\end{center}
\caption{
Diagrams contributing to the decay width of $p$-Ps at $O(\alpha^2)$
and their reduction to simpler master integrals. 
Dashed and solid lines represent massless and masive lines, respectively.}
\end{figure}

  This quantity recieves contributions from two-loop diagrams of $e^+e^-$
annihilation into two photons in threshold kinematics. However, the
generic planar and non-planar diagrams (see Fig.~2, upper row) can be
reduced via integration by parts to simpler integrals (Fig.~2, middle row).
These, in turn, as we shall see, contain constants that are related to the
sunset diagram (Fig.~2, bottom row) at very special kinematics, namely when
the external momentum $q$ is restricted by $q^2=-m^2$. The sunset diagrams with
such kinematics have been considered in great detail in \cite{I1}.
In particular the result for the sunset is expressed in terms of special sums
osf elliptic nature,
\begin{eqnarray}
\sum_{n=1}^{\infty} (-1)^n
    \frac{\left(2n \atop n\right)}{\left(4n \atop 2n\right)}
    \left\{  \phi,\, \frac{\phi}{n},\, \frac{1}{n^2}
       \right\} \,,
\label{aa}
\end{eqnarray}
which we can call $a_\phi,\, a_{\phi 1}$ and $a_2$, respectively, and other sums
\begin{eqnarray}
\sum_{n=1}^{\infty} 
    \frac{(-16)^n}{\left(2n \atop n\right)\left(4n \atop 2n\right)}
    \left\{  1,\, \frac{1}{n}
       \right\} \,,
\label{bb}
\end{eqnarray}
which we call $b_0$ and $b_1$. In (\ref{aa}), $\phi$ stands for
$$
 \phi = S_1(n-1) - 3S_1(2n-1)+2S_1(4n-1) \,,
$$
with $S_a(n)=\sum_{j=1}^n 1/j^a$ being a harmonic sum.

\begin{figure}[ht]
\begin{center}
\includegraphics[width=0.20\textwidth]{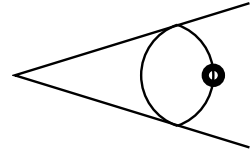}
\end{center}
\caption{
Example of vertex diagram $J$, contributing to the decay width of $p$-Ps.
All lines have mass $m$. The dot on a line means the square of the
propagator.
}
\end{figure}

  Starting from (\ref{aa}) and (\ref{bb}), one can construct sums of higher weights, 
e.g. $a_3,\, a_{\phi2},\, b_3$, etc. With such constructed sums, we evaluate
more complicated diagrams, including vertexes and boxes. We illustrate it
evaluating diagram $J$ shown in Fig.~3. The resul is
\begin{equation}
  J = \frac{9}{16} \zeta(3) - \frac{1}{8} a_3 - \frac{1}{8} a_{\phi2}
      - \frac{1}{32} b_3
\label{J}
\end{equation}
and a similar result follows for the box diagrams of Fig.~2. Formula (\ref{J})
shows the deep relation of the vertex diagram with the sunset diagram
(in fact such relation follows from the differential equations).

  Concluding this section we want to mention that there are relations
between the above sums and also their relation to the elliptic integrals
has been found in \cite{I1}.

\section{Conclusions}

  Thus, we established the analytical structure of the results for the nex unknown
corrections both for ortho- and parapositronium lifetimes. We found that new
constants, that are not related to the Euler--Zagier sums appear in both cases,

  We are grateful to G. S. Adkins for providing us with the computer code
employed for the numerical analysis in \cite{Adkins:2005eg} and M. Yu. Kalmykov
for the fruitful discussions.

\end{document}